\documentclass[aps,twocolumn,showpacs,preprintnumbers,amsmath,amssymb]{revtex4}
\usepackage{graphicx}
\usepackage[]{caption}
\usepackage{amsmath}
\usepackage{amssymb}

\usepackage{graphicx}
\usepackage{dcolumn}
\usepackage{bm}
\def\be{\begin{equation}}
\def\ee{\end{equation}}
\def\bea{\begin{eqnarray}}
\def\eea{\end{eqnarray}}

\begin{document}

\title{Cosmological Perturbations in Non-Commutative Inflation}

\author{Seoktae Koh} \email[email: ]{skoh@hep.physics.mcgill.ca} 
\author{Robert H. Brandenberger} \email[email: ]{rhb@hep.physics.mcgill.ca}

\address{Physics Department, McGill University, Montr\'eal, Q.C., 
H3A 2T8, Canada}


\begin{abstract}

We compute the spectrum of cosmological perturbations in a
scenario in which inflation is driven by radiation in
a non-commutative space-time. In this scenario, the
non-commutativity of space and time
leads to a modified dispersion relation
for radiation with two branches, which allows for
inflation. The initial conditions for the cosmological
fluctuations are thermal. This is to be
contrasted with the situation
in models of inflation in which the accelerated
expansion of space is driven by the potential energy
of a scalar field, and in which the fluctuations are
of quantum vacuum type. We find that, in the limit
that the expansion of space is almost exponential, the
spectrum of fluctuations is scale-invariant with a slight
red tilt. The magnitude of the tilt is
different from what is obtained in a usual
inflationary model with the same expansion rate
during the period of inflation. The amplitude also
differs, and can easily be adjusted to agree
with observations.  

\end{abstract}

\maketitle

\section{Introduction}

Various approaches to quantum gravity indicate that the effective field
theory which emerges below the cutoff scale will be non-commutative.
For example, in the context of string theory, one of the reasons for 
non-commutativity is that the basic objects which underlie the quantum
theory are strings rather than point particles. A concrete formulation
of this non-commutativity comes from the ``stringy space-time uncertainty
relation'' of \cite{Yoneya,Li}. In the matrix theory approach to 
non-perturbative string theory, spatial coordinates arise in a certain
limit from matrices which are non-commuting \cite{Jevicki,BFSS,IKKT}.
This leads to space-space non-commutativity. 

The most promising arena to probe the
fundamental non-commutativity of space-time on microscopic scales is
cosmology. The reason is that, according to our present understanding,
the seeds for the currently observed structure of the universe on large
scales were laid down in the very early universe when the length
scales which we probe today in cosmology were microscopic and thus
subject to the modifications in the dynamics which arise due to the
non-commutativity.

The inflationary universe scenario is the current paradigm for early
universe cosmology \cite{Guth} (see also 
\cite{Starob1,Sato,Brout}). According
to this scenario, the universe experiences a period of accelerated
expansion in its early stages. Often, the expansion is close
to exponential, or it is described by the scale factor $a(t)$
expanding as a large power $p > 1$ of time t. This accelerated 
expansion leads to the possibility of a causal structure formation
scenario, as first discussed in \cite{Mukh1} (see also 
\cite{Starob2,Press,Sato}). Figure 1 provides a
space-time sketch illustrating the basic idea. During
the period of acceleration, fixed comoving scales are stretched
beyond the Hubble radius $H^{-1}$. Fluctuations on these scales
can become the seeds for the presently observed large-scale 
structure and the anisotropies in the cosmic microwave background.

By the same argument, however, we can argue \cite{RHBrev1}
that fluctuations
on currently observed scales had a wavelength during the early stages
of inflation which was so small that effects of Planck-scale
physics will be operative and leave an imprint on the initial
evolution of the fluctuations, effects which are preserved until
late times by the linear evolution of the fluctuations. 
Initial studies of effects of trans-Planckian physics on current
observations in the context of inflationary cosmology made use
of ad hoc modified dispersion relations for the 
fluctuation modes \cite{Martin}. These initial studies were
followed by specific investigations of the effects of
space-space \cite{SSUR} and space-time \cite{Ho} non-commutativity.
All of the works referenced in this paragraph were based on
the current models of inflation in which it is assumed that
inflation is driven by the potential energy of a scalar matter
field. In this context, any pre-existing classical fluctuations
at the beginning of the phase of inflation are red-shifted, leaving
behind a quantum vacuum. Hence, it was postulated that the 
currently observed fluctuations are the result of the non-trivial
evolution of quantum vacuum perturbations (see e.g. \cite{MFB}
for a comprehensive review of the theory of cosmological perturbations,
and \cite{RHBrev2} for a pedagogical introduction).

Non-commutativity of space-time or space-space leads to
deformed dispersion relations (see e.g. \cite{Kempf}).
Whereas space-space non-commutativity introduces 
anisotropy into the dispersion relations, space-time
non-commutativity leads to dispersion relations which
preserve isotropy. In \cite{Joao1} 
a class of deformed dispersion relation for ordinary radiation 
resulting from space-time non-commutativity was studied.
This dispersion relation has a maximal
momentum, and it has two branches (two values
of frequency for fixed momentum).
In \cite{Joao2} it was pointed out that this dispersion
relation makes it possible to
obtain a period of inflation driven by non-commutative radiation.
Typically, one obtains power-law inflation, inflation
driven by the radiation itself. In \cite{Joao2},
an order of magnitude estimate of the amplitude of the resulting
cosmological fluctuations was given. In this paper, we present
a detailed study of the spectrum of cosmological perturbations
in this model of non-commutative inflation. In contrast to
the conventional models of inflation where the fluctuations
are of quantum vacuum origin, in our model they are due to
thermal fluctuations \footnote{See also \cite{Levon} for
a study of thermal fluctuations in the context of general
cosmological models, and \cite{NBV,BNPV2} for a new structure
formation scenario based on string gas cosmology which makes
use of string thermodynamic fluctuations.}. In this aspect,
our model has analogies with the ``warm inflation'' scenario
\cite{warm} in which the fluctuations are also of thermal
origin. 

The following Section provides a brief review of the 
non-commutative inflation scenario of \cite{Joao1,Joao2}.
Section 3 gives a summary of the key equations which
describe the evolution of cosmological fluctuations which we
will make use of. In Section 4, we present the computation
of the spectrum of scalar metric perturbations, based on
thermal initial values for the fluctuations. We conclude
with a brief discussion.

\begin{figure}
\includegraphics[height=8cm]{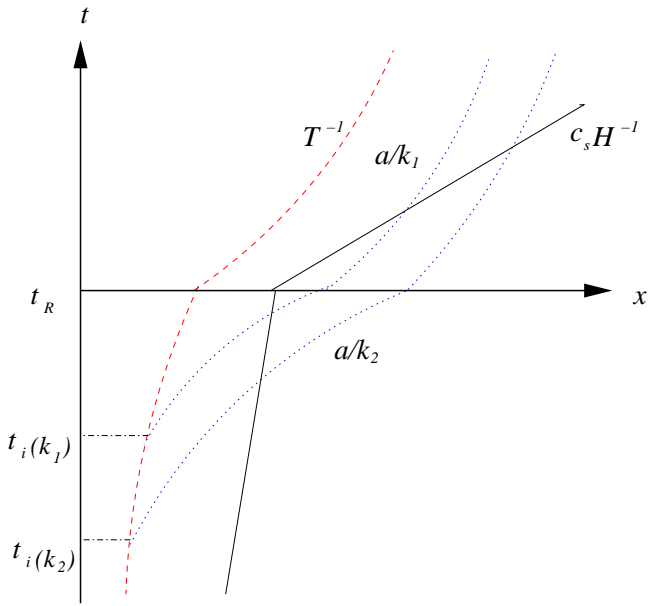}
\caption{Space-time sketch depicting the expansion of fixed
comoving scales in inflationary cosmology. The horizontal axis
represents the physical distance, the vertical axis time. The
time $t_R$ denotes the transition time between the inflationary
phase and the radiation phase of standard cosmology. The curves
corresponding to the physical wavelengths of two comoving scales
with comoving wavenumbers $k_1$ and $k_2$ are labelled by their
wavenumbers. The solid
line indicates the sound horizon (up to an irrelevant constant
equal to the Hubble radius), the dashed line is the thermal
correlation length. Note that the slopes of these two curves are
different.}
\label{fig:1}
\end{figure}
 
\section{Non-Commutative Inflation}

In the case of massless particles, the non-commutativity 
of space-time leads to modifications of the
usual linear dispersion relation which take the form
\be \label{disprel1}
E^2  - p^2 c^2 f(E)^2 \, = \, 0 \,
\ee
with
\be \label{disprel2}
f(E) \, = \, 1 + (\lambda E)^{\alpha} \, .
\ee
In the above, $p$ and $E$ denote momentum and energy,
respectively, $\alpha$ is a positive constant, and
$\lambda$ is a length scale whose physical meaning is
the following: solving (\ref{disprel2}) for the
momentum $p$, we see immediately that for $\alpha \ge 1$
there is a maximal momentum $p_{max}$ whose value is
determined by $\lambda$. For example, for $\alpha = 1$
the maximal momentum is $p_{max} = 1/(c \lambda)$.

For $\alpha > 1$, there are two values of the energy $E$
for any momentum - the dispersion relation has two
branches. On the upper branch, the energy increases as
the momentum decreases, which is what happens when the
universe expands. It is this property which is crucial
in order to realize that modified dispersion relations
with $\alpha > 1$ can, at least for a certain range
of values of $\alpha$, lead to inflation driven by
ordinary radiation.

The modification of the dispersion relation leads to a
deformed thermal spectrum \cite{Joao1} due to the 
changing  density of states
\be
\rho(E) \, = \, \frac{1}{\pi^2 \hbar^3 c^3}\frac{E^3}{e^{\beta E}-1}
\frac{1}{f^3}\left|1-\frac{f^{\prime}E}{f}\right| \, ,
\ee
where $\rho(E)$ denotes the energy density (per unit $E$)
and $\beta$ is the inverse temperature \footnote{The overall
energy density is denoted by $\rho$.}.
In the high energy limit, $f^{\prime}E/f \simeq \alpha$. The
high energy Stephan-Boltzmann equation takes the form $\rho 
\propto T^{\gamma}$  with $1<\gamma <4$ where $\gamma =4$ for 
$\alpha=0$ and $\gamma=1$ for $\alpha \geq 1$.

The equation of state which follows from the modified dispersion
relation is given by
\be
{\cal P} \, = \, \frac{1}{3}\int \frac{\rho(E) dE}{1-\frac{f^{\prime}E}{f}} \, ,
\ee
where ${\cal P}$ stands for the pressure.
This is approximated in the high energy limit by
\be
w(\rho \rightarrow \infty) \, \simeq \, \frac{1}{3(1-\alpha)} \, ,
\ee
where $w = {\cal P} / \rho$ is the usual equation of state parameter.
For $\alpha > 1$, the equation of state parameter $w$ takes on negative values and
it is possible to have inflationary expansion with $-1 \leq w
< -1/3$ for temperatures $T\gg 1/\lambda$.

The Friedmann and energy conservation equations are 
\begin{eqnarray}
\left(\frac{\dot{a}}{a}\right)^2 \, &=& \, \frac{8\pi}{3M_p}\rho, \\
\dot{\rho} \, &=& \, - 3\frac{\dot{a}}{a}(1+w)\rho \nonumber
\end{eqnarray}
and its solutions are given for a constant equation of state parameter by
\begin{eqnarray}
\rho \, &\propto& \, a^{-3(1+w)} \, \sim \,
a^{-\frac{4-3\alpha}{1-\alpha}}  \nonumber \\
 a(t) \, &\propto& \, t^{\frac{2}{3(1+w)}} \, 
\propto \, \eta^{\frac{2}{1+3w}}
\label{bg_sol}
\end{eqnarray}
These show power law inflation ($-1 \le w <-1/3$) for values
of $\alpha$ in the range $4/3 \leq \alpha <2$ \footnote{The full
numerical solution of \cite{Joao2} shows that this range of
$\alpha$ values shrinks at the upper end compared to what is
obtained using the above approximate analysis.}.
For $\alpha \simeq 4/3$ we have almost exponential inflation.
Note that for $1 < \alpha < 4/3$, non-commutative radiation at $T \gg 1/\lambda$
behaves like phantom matter ($w <-1$). Inflation ends at 
time $t_R$ when there are no more excited states on the top
branch of the dispersion relation. 

\section{Key tools of the theory of cosmological perturbations}

In this section we will list the key equations from 
the theory of cosmological perturbations which  will be used.
We work in longitudinal gauge in which the metric takes the form
\be
ds^2 \, = \, a^2 [ -(1+2\Phi)d\eta^2 +(1-2\Psi)\gamma_{ij}dx^i dx^j] \, ,
\ee
where $a(\eta)$ is the scale factor, $\eta$ being conformal time, and
the functions $\Phi$ and $\Psi$ describe the scalar metric
fluctuations and thus depend on space and time. At the linearized level, 
the perturbed Einstein equations can be written as \cite{MFB}
\begin{eqnarray}
& &3\mathcal{H}^2 \Phi+3\mathcal{H}\Psi^{\prime}-\nabla^2 \Psi 
= -4\pi G a^2 \delta {T^0}_0,
\label{pert_g00} \\
& &(\mathcal{H}\Phi+\Psi^{\prime})_{,i} = 4\pi G a^2 \delta {T^0}_i,
\label{pert_g0i}\\ 
& &[(2\mathcal{H}^{\prime}+\mathcal{H})\Phi+\mathcal{H}\Phi^{\prime}
+\Psi^{\prime\prime}+2\mathcal{H}\Psi^{\prime}]
+\frac{1}{2}\nabla^2(\Phi-\Psi) \nonumber \\
& & \,\,\,\,\, = -4\pi G a^2 \delta {T^i}_i, 
\label{pert_gij}\\
& &{(\Phi - \Psi)^{,i}}_{,j} = -4\pi G a^2 \delta {T^i}_j, ~~(i\neq j) \, .
\end{eqnarray}
In the above, $G$ denotes
Newton's gravitational constant and ${\cal H} = a^{\prime}/a$, a prime 
denoting the derivative with respect to $\eta$.

Since there is no anisotropic stress, one degree of freedom for
scalar metric fluctuations disappears and we can set $\Phi = \Psi$ 
\cite{MFB}. Eqs. (\ref{pert_g00}) and (\ref{pert_gij}) can then 
be  combined to give
\begin{eqnarray}
\Phi^{\prime\prime} &+& 3\mathcal{H}(1+c_s^2)\Phi^{\prime}
+ c_s^2 k^2 \Phi \nonumber \\
&+& (2\mathcal{H}^{\prime} + (1+3c_s^2)\mathcal{H}^2)\Phi \, = \, 0 \, ,
\label{2ndeq}
\end{eqnarray}
where we have used $\delta {\cal P} = c_s^2 \delta \rho$. As is well known,
as long as the equation of state is not changing, then
on scales larger than the sound horizon one of the solutions of the
above equation tends to a constant, the other is decaying. Thus, to
know the power spectrum of $\Phi$ at late times on scales larger 
than the sound horizon,
it is sufficient to evaluate the amplitude of the $k$'th mode of $\Phi$
when the wavelength is equal to the sound horizon.

Introducing a variable $v$ (in terms of which the action for
cosmological perturbations has canonical kinetic term \cite{Mukh2,Sasaki})
via
\be \label{vvar}
\Phi \, = \, 
4\pi G\sqrt{\rho+{\cal P}}\frac{z}{k^2 c_s}\left(\frac{v}{z}\right)^{\prime} \, ,
\ee
the equation of motion for scalar metric perturbations takes the simple form
\be
v^{\prime\prime} + \left(k^2 c_s^2 - \frac{z^{\prime\prime}}{z}\right)v \,
= \, 0 \, ,
\label{mode_eq}
\ee
where
\be
z \, = \, \frac{a\sqrt{\rho+{\cal P}}}{\mathcal{H}c_s} \, .
\label{z}
\ee
If the equation of state does not change in time, then the
variable $z$ is proportional to the scale factor $a$. 

If the equation of state changes, as it does at the end of
the phase of inflation, $\Phi$ is not constant. Fortunately,
for purely adiabatic perturbations there is a quantity which
is conserved on scales larger than the sound horizon. This
quantity is \cite{BST,BK}
\be \label{zetaeq}
\zeta \, \equiv \, \Phi + {2 \over 3} 
{{\bigl(H^{-1} {\dot \Phi} + \Phi \bigr)} \over { 1 + w}} \, .
\ee
Considering a transition between the period of radiation-driven
inflation to the post-inflationary phase of regular radiation
with $w = 1/3$, the conservation of $\zeta$ implies
\be \label{factor1}
\Phi_f \, = \, \Bigl( {2 \over 3} + {4 \over {3(1 + w)}} \Bigr) \Phi_i
\, ,
\ee
where $\Phi_i$ is the super-horizon value of $\Phi$ during
the initial period and $\Phi_f$ is its final value during
the radiation phase of standard cosmology. Thus, the final
spectrum of $\Phi$ is related to the initial spectrum by
a scale-independent multiplicative factor ${\cal F}$, which
in the case of observationally viable backgrounds for which $w$ is close
to $-1$ is approximately
\be \label{factor2}
{\cal F} \, \simeq \, {4 \over {3(1 + w)}} \, .
\ee 

In the following we will compute the spectrum of cosmological
perturbations assuming that their origin is thermal. The
correct variable to consider in this context is $v$. The first
justification is that it is this variable which enters the
action for cosmological perturbations in the canonical way, and
in terms of which concepts like particle number are well-defined.
A second reason is that on sub-Hubble scales, the variable $v$
for scalar field matter reduces to the matter field fluctuation
(multiplied by the scale factor). Hence, the concept
of ``number of quanta of $v$'' reduces to the concept of the number
of field fluctuation quanta \footnote{Note also
that $v$ is related to the curvature perturbation $\mathcal{R}$ in
comoving gauge as $v = z\mathcal{R}$.}.

We will assume that the evolution equation for $v$ is not
modified by the underlying space-time non-commutativity. 
This is well justified provided that the
wavelengths we are interested in are much larger than the minimal
length $\lambda$. On scales larger than the Hubble radius this
is always the case. If we consider temperatures $T$ smaller than
$\lambda^{-1}$, then we are safe for both sets of initial conditions
we consider (see below).

\section{Density perturbation in non-commutative inflation}

The key question which has to be addressed is on what scale
the initial conditions should be imposed. On a fixed comoving
scale, it should be the time after which that scale ceases to be
in thermal equilibrium. There are two possibilities: the Hubble
scale (beyond which causality prohibits local causal interactions
\cite{Traschen}) or else the thermal correlation length
$T^{-1}$. In the following, we will adopt the latter
prescription. This means that for a fixed scale $k$, we
will fix the fluctuations of the canonical variable $v$ to
have a thermal equilibrium number at the time $\eta_i(k)$
given by
\be
T^{-1}(\eta_i(k)) \, = \, {{a(\eta_i(k))} / k} \, . 
\ee

The starting point of the analysis is the quantization of the linear
cosmological perturbation variable $v$. This variable is
expanded into creation and annihilation operators $a^{\dag}_k$
and $a_k$ in the standard way. Thus, for a fixed $k$, the state
$|\theta\rangle$ at time $\eta_i(k)$ is determined by
\be \label{thr1}
\langle \theta | \hat{N}_k |\theta \rangle 
\, = \, \langle \theta| a^{\dag}_k a_k |\theta \rangle 
\, = \, n_k \, ,
\ee
where $\hat{N}_k = a_k^{\dag}a_k$ is the number operator, and 
the expectation value $n_k$ is set by the Bose-Einstein distribution
\be \label{thr2}
n_k \, = \, \frac{1}{e^{\frac{E}{aT}}-1} \, .
\ee

In order to connect to observations, we need to compute the
power spectrum of the relativistic potential $\Phi$. Making use
of (\ref{vvar}), this power spectrum is given by
\begin{eqnarray}
P_{\Phi}(k) & \equiv & 
\frac{k^3}{2\pi^2}\langle \theta| \Phi_k^{\ast}\Phi_k |\theta\rangle 
\nonumber \\
& = & \frac{1}{M_p^4}\frac{8(\rho+p)z^2}{k c_s^2}
\left\langle \theta\left|\left(\frac{v_k^{\ast}}{z}\right)^{\prime}
\left(\frac{v_k}{z}\right)^{\prime}\right|\theta\right\rangle
\label{ps}
\end{eqnarray}

It is convenient to use the Heisenberg picture to quantize the system
especially when the vacuum is time-dependent. In this picture, 
the operators
evolve with time, but the states do not. The operators $v_k$ and their conjugate momenta $\pi_k$ which are given by
\be \label{piex}
\pi_k \, = \, v_k^{\prime} - \frac{z^{\prime}}{z}v_k
\ee
can be written in terms of creation and annihilation operators as
\begin{eqnarray}
v_k(\eta) \, &=& \, 
\frac{1}{\sqrt{2kc_s}} [a_k(\eta) + a_{-k}^{\dag}(\eta)], \nonumber \\
\pi_k(\eta) \, &=& \, 
- i\sqrt{\frac{kc_s}{2}}[a_k(\eta) + a_{-k}^{\dag}(\eta)] \, ,
\label{mode_ex}
\end{eqnarray}
where we have used
\be
a_k^{\prime} \, = \, i[a_k, H_k]
\ee
and similarly for $a^{\dag}_k$, and the
Hamiltonian $H_k$ is given by
\be
H_k \, = \, \int d^3k \frac{1}{2}\left[
k(a_k a^{\dag}_k+a^{\dag}_k a_k)+i\frac{z_k^{\prime}}{z_k}
(a^{\dag}_k a^{\dag}_{-k}-a_k a_{-k})\right] \, .
\ee

The annihilation and creation operators at time $\eta$ are related to
those at the initial time $\eta_i$ by the Bogoliubov transformation
\begin{eqnarray}
a_k(\eta) \, &=& \, 
\alpha_k(\eta) a_k(\eta_i) + \beta_k(\eta)a_{-k}^{\dag}(\eta_i), \\
a^{\dag}_{-k}(\eta) \, &=& \, 
\alpha_k^{\ast}(\eta)a_{-k}^{\dag}(\eta_i)
+ \beta_k^{\ast}(\eta)a_k(\eta_i) \, , \nonumber
\label{bog_transf}
\end{eqnarray}
where $\alpha_k$ and $\beta_k$ are the Bogoliubov coefficients
which satisfy the relation $|\alpha|^2 -|\beta|^2 \, = \, 1$. 
Substituting these relations into Eqs. (\ref{mode_ex}) yields
\begin{eqnarray}
v_k (\eta) \, &=& \, 
f_k(\eta)a_k(\eta_i) + f_k^{\ast}(\eta) a_{-k}^{\dag}(\eta_i), \\
\pi_k (\eta) \, &=& \, 
-i[g_k(\eta)a_k(\eta_i) + g_k^{\ast}(\eta)a_{-k}^{\dag}(\eta_i)] \, ,
\nonumber
\end{eqnarray}
where
\begin{eqnarray}
f_k(\eta) \, &=& \,  
\frac{1}{\sqrt{2k c_s}}(\alpha_k(\eta) + \beta_k^{\ast}(\eta)), 
\nonumber  \\
g_k(\eta) \, &=& \, 
\sqrt{\frac{k c_s}{2}}(\alpha_k(\eta)-\beta_k^{\ast}(\eta)) \nonumber \\
&=& i\left(f_k^{\prime} - \frac{z^{\prime}}{z}f_k\right),
\label{mode_bogcof}
\end{eqnarray}
and $f_k$ is a solution of the mode equation (\ref{mode_eq}).

From Eqs. (\ref{bg_sol}) and (\ref{z}) it follows that
\be
\frac{z^{\prime\prime}}{z} \, = \, \frac{a^{\prime\prime}}{a}
\, = \, \frac{\nu^2 -1/4}{\eta^2} \, ,
\ee
where
\be
\nu^2 - \frac{1}{4} \, = \, \frac{2(1-3w)}{(1+3w)^2} \, .
\ee
The mode equation then has the following solution
\be
f_k(\eta) \, = \, A_k\sqrt{-\eta}J_{\nu}(-k c_s \eta)
+ B_k \sqrt{-\eta}Y_{\nu}(-k c_s \eta)
\ee
where $J_{\nu}$ and $Y_{\nu}$ are Bessel functions of the
first and second kind, respectively. 
From Eq.(\ref{mode_bogcof}), we then obtain
\be \label{gexpr}
g_k(\eta) \, = \,  -ikc_s\sqrt{-\eta}(A_k J_{\nu-1}(-k c_s\eta)
+B_k Y_{\nu-1}(-k c_s \eta)) \, .
\ee

Note that for $\alpha= 4/3~ (w= -1)$ we have 
$\nu^2 -1/4 = 2$, and thus the mode functions $f_k$ can be written as
\begin{eqnarray}
f_k(\eta) \, &=& \, \frac{A_k}{\sqrt{2k c_s}}e^{-ikc_s \eta}
\left(1-\frac{i}{kc_s \eta}\right) \nonumber \\
&+& \frac{B_k}{\sqrt{2k c_s}}
e^{ikc_s\eta}\left(1+\frac{i}{kc_s \eta}\right) \, .
\end{eqnarray}
and 
\be
g_k(\eta) \, = \,  A_k\sqrt{\frac{kc_s}{2}}e^{-ikc_s \eta}
-B_k\sqrt{\frac{kc_s}{2}}e^{ikc_s\eta} \, .
\ee

We can obtain the results in terms of $\alpha_k$ and $\beta_k$ using
Eq.(\ref{mode_bogcof}):
\begin{eqnarray}
\alpha_k \, &=& \, 
\sqrt{\frac{-kc_s\eta}{2}}[A_k(J_{\nu}(-kc_s\eta)
-iJ_{\nu-1}(-kc_s\eta)) \, \nonumber \\ 
& & + B_k(Y_{\nu}(-kc_s)-iY_{\nu-1}(-kc_s\eta))], \nonumber \\
\beta_k^{\ast} \, &=& \, 
\sqrt{\frac{-kc_s\eta}{2}}[A_k(J_{\nu}(-kc_s\eta)
+i J_{\nu-1}(-kc_s\eta)) \nonumber \\
& & + B_k(Y_{\nu}(-kc_s\eta)+iY_{\nu-1}(-kc_s\eta))]
\end{eqnarray}
The normalization condition for $\alpha_k$ and $\beta_k$ implies that
\be
A_k B_k^{\ast}-A_k^{\ast}B_k \, = \, -i\frac{\pi}{2} \, .
\ee
Since $\beta_k^{\ast}(\eta_i)= 0$ 
at the initial time $\eta_i$, we obtain
\be
A_k = -\frac{Y_{\nu}(-kc_s\eta_i)+iY_{\nu-1}(-kc_s\eta_i)}
{J_{\nu}(-kc_s\eta_i)+iJ_{\nu-1}(-kc_s \eta_i)}B_k
\label{cof1}
\ee
from which it follows that
\begin{eqnarray}
|A_k|^2 \, &=& \, -\frac{\pi^2}{8}kc_s\eta_i(Y_{\nu}^2(-kc_s\eta_i)
+Y_{\nu-1}^2(-kc_s\eta_i)), \nonumber \\
|B_k|^2 \, &=& \, -\frac{\pi^2}{8}kc_s\eta_i(J_{\nu}^2(-kc_s\eta_i)
+J^2_{\nu-1}(-kc_s\eta_i)), \nonumber \\
A_k B_k^{\ast} \, &=& \, -i\frac{\pi}{4}+\frac{\pi^2}{8}k c_s \eta_i
(J_{\nu}(-kc_s\eta_i)Y_{\nu}(-kc_s\eta_i) \nonumber \\
& & + J_{\nu-1}(-k c_s\eta_i)Y_{\nu-1}(-k c_s\eta_i)) \label{cof2} .
\end{eqnarray}

The power spectrum of the perturbation variable $\Phi$ at a late
time $\eta$ resulting from taking thermal initial conditions at
the times $\eta_i(k)$ can now be calculated as follows:
\begin{eqnarray}
& & P_{\Phi}(k, \eta) \, = \, \frac{1}{M_p^4}\frac{8(\rho+p)z^2}{k c_s^2}
\left\langle \theta\left|\left(\frac{v_k^{\ast}}{z}\right)^{\prime}
\left(\frac{v_k}{z}\right)^{\prime}\right|\theta\right\rangle \nonumber \\
&=& \, \frac{1}{M_p^4}\frac{8(\rho+p)}{kc_s^2}
\langle \theta|\pi_k^{\ast}\pi_k|\theta\rangle \nonumber \\
&=& \, \frac{1}{M_p^4}\frac{8(\rho+p)}{kc_s^2}|g_k|^2
(2n_k(\eta_i)+1) \\
&=& \, \frac{1}{M_p^4}\frac{8(\rho+p)}{k c_s^2}|g_k|^2 \coth
\left(\frac{kc_s}{a_i T_i}\right) \nonumber \\
&=&\frac{8(\rho+p)k\eta}{M_p^4}\coth
\left(\frac{kc_s}{a_i T_i}\right) \nonumber \\
& & [|A_k|^2 J_{\nu-1}^2(-kc_s\eta)
+ |B_k|^2 Y_{\nu-1}^2(-k c_s \eta) \nonumber \\
& &+ (A_k B_k^{\ast} + A_k^{\ast} B_k)
J_{\nu-1}(-kc_s \eta) Y_{\nu-1}(-k c_s\eta)] \nonumber
\label{ps_phi}
\end{eqnarray}
where $a_i = a(\eta_i(k))$ and $T_i = T(\eta_i(k))$, and the
functions are evaluated at the time $\eta$ unless indicated otherwise. 
To go from the first
to the second line, we have used (\ref{piex}), to go from the second to the
third line (\ref{mode_ex}) and (\ref{thr1}),
from the third to the fourth line (\ref{thr2}), and in the last step
(\ref{gexpr}). Inserting the values of the coefficients from (\ref{cof2})
we obtain
\begin{eqnarray}
P_{\Phi}(k) \, &=& \, \frac{\pi^2(\rho+p)k^2c_s^2\eta\eta_i}{M_p^4 c_s}
\coth\left(\frac{kc_s}{a_i T_i}\right) \nonumber \\
& & [-(Y_{\nu}^2(-kc_s\eta_i) + Y_{\nu-1}^2(-kc_s\eta_i))J_{\nu-1}^2(-kc_s\eta) \nonumber \\
& & \, -(J_{\nu}^2(-kc_s\eta_i) + J^2_{\nu-1}(-kc_s\eta_i)) Y_{\nu-1}^2(-k c_s \eta) \nonumber \\
& & \, + 2(J_{\nu}(-kc_s\eta_i)Y_{\nu}(-kc_s\eta_i) \nonumber \\
& & \, + J_{\nu-1}(-k c_s\eta_i)Y_{\nu-1}(-k c_s\eta_i)) \nonumber \\
& & J_{\nu-1}(-kc_s \eta) Y_{\nu-1}(-k c_s\eta)] \, .
\end{eqnarray}
Since $\Phi$ is constant on super-sound horizon scales, it is sufficient
to evaluate $P_{\Phi}(k)$ at sound-horizon crossing $\eta_h(k)$ and to
take $\Phi$ to be constant after that.

If we impose thermal initial conditions at sound horizon crossing,
then $\eta_i(k) = \eta_h(k)$ and $kc_s\eta_i = kc_s\eta_h =1$. Thus,
up to constants of order unity, the amplitude of the power spectrum is
given by
\be
P_{\Phi}(k) \, \sim \, {\cal F} \frac{\pi^2 (1 + w) \rho(\eta_h(k))}{M_p^4} \, ,
\ee
where ${\cal F}$ is the factor (\ref{factor2}) 
relating the amplitude of the
power spectrum during the inflationary phase to that in the
post-inflationary radiation phase of Standard Cosmology. The 
spectral index is
\be
n-1 \, = \, \frac{d\ln P_{\Phi}(k)}{d\ln k} \,
 = \, \frac{d\ln \rho(k)}{d\ln k} \,
= \, \frac{6(1+w)}{1+3w}.
\ee
This gives a scale invariant spectrum in the limit $w\rightarrow -1$
with a slight red tilt. The tilt is the same as what is obtained
in scalar field-driven commutative inflation with the same expansion
rate.

This choice for initial conditions is, however, not natural because thermal
equilibrium will not be maintained on scales larger than the thermal 
correlation length.

Thus, we go on to consider the more natural choice $\eta_i <\eta_h$ with
$\eta_i(k)$ given by when the scale $k$ equals the thermal correlation
length. Since $kc_s =a_i T_i = a(\eta_h)H(\eta_h)$, the amplitude of
the power spectrum then becomes
\be
P_{\Phi}(k) \, \sim \, {\cal F} \frac{(1 + w)\rho}{M_p^4}\frac{T_i}{H_i}
\ee
where we have used $\eta_i \simeq a_i^{-1}H_i^{-1}$. Note that the
amplitude obtained with this prescription for setting initial conditions
is larger than what was obtained using the previous recipe, the reason
being that in order to maintain thermal equilibrium number of modes
while the physical momentum is decreasing, the mode occupation number
needs to decrease via interactions. This does not happen in the second
prescription. Note also that the amplitude of the spectrum obtained
with this second prescription agrees with the heuristic estimate of
\cite{Joao2}.

If we use the following relations
\begin{eqnarray}
kc_s \, &=& \, a_i T_i \, \sim \, a_i \rho_i \, 
\sim \, a_i^{-2-3w}, \nonumber \\
a_i \, &\sim& \, \eta_i^{\frac{2}{1+3w}},
\quad \eta_i \, \sim \, k^{-\frac{1+3w}{2(2+3w)}}, 
\end{eqnarray}
then the spectral index of the cosmological fluctuations becomes
\begin{eqnarray}
n-1 \, &=& \, \frac{d\ln P_{\Phi}(k)}{d\ln k} \, 
\nonumber \\
& = & \, \frac{9(1+w)(3+5w)}{2(1+3w)(2+3w)} \, .
\end{eqnarray}
This also gives a scale invariant spectrum in the limit
$w \rightarrow -1$ with a slightly red tilt. Note, however,
that the magnitude of the tilt differs from what is obtained
in scalar field-driven inflation with the same power law
expansion of the scale factor.
Note that we can also get a blue tilted spectrum 
in the limit $w \rightarrow -\frac{3}{5}$ for
the range $-2/3<w <-3/5$.

\section{Discussion}

We have computed the spectrum of cosmological perturbations in
the non-commutative inflation model of \cite{Joao2}, in which
the accelerated expansion of space is generated by the modified
dispersion relation of ordinary radiation. The dispersion relation
has two branches, i.e. two values of the frequency for every
wavenumber. The upper branch of the dispersion relation 
has increasing frequency as $k$ decreases, this being the key
property which leads to inflation. In the context of this model 
of inflation, the cosmological fluctuations are of thermal origin
- they are the thermal equilibrium fluctuations of the same
radiation fluid which generates inflation. 

We discuss two prescriptions for setting up initial conditions.
In the first, we set up the cosmological fluctuations mode by mode
with thermal occupation numbers at the times when the modes exit
the Hubble radius. In the second prescription, the modes are
initialized with thermal occupation numbers at the times when their
wavelengths equal the thermal correlation length.
We believe that this second prescription is the more realistic one,
since on scales larger than the thermal correlation length the
interactions are not likely able to maintain thermal occupation numbers.

Both prescriptions for initial conditions lead to a spectrum of
fluctuations which is scale-invariant in the limit in which the
expansion of space becomes exponential. For nearly exponential
expansion the spectrum has a slight red tilt. We have computed the amount
of the red tilt. For the choice of initial conditions which
we believe are realistic in the context of model of inflation, the tilt 
has a different numerical value compared to the result for vacuum initial conditions
in a scalar field-driven commutative inflation model with the same
expansion rate. The amplitude of the spectrum is
different from what is obtained in regular inflation models with
the same background expansion history. 

Our model is not the first inflationary model in which thermal rather
than quantum vacuum initial conditions are the source of the
inhomogeneities. The warm inflation \cite{warm} scenario also has
dominant thermal fluctuations. However, in warm inflation the acceleration
of the background is generated by a commutative scalar field
rather than by non-commutative radiation. Thermal rather than
vacuum fluctuations also play a crucial role in the recently
discovered string gas structure formation scenario \cite{NBV,BNPV2}.
In this scenario, the fluctuations are of string thermodynamic origin
rather than given by point particle thermodynamics.

The possibility of thermal fluctuations as the source of an almost
scale-invariant spectrum has also been studied in \cite{Levon}.
However, in that work a regular dispersion relation for
radiation was assumed. Hence, no accelerated expansion resulted,
and it was not possible to generate a nearly scale-invariant
spectrum in a cosmological background in which the universe was
always expanding.

Our work yields a confirmation of the fact that in inflationary
cosmology, the accelerated expansion of space makes it possible
to probe Planck-scale physics in current observations via the
induced signals in the spectrum of cosmological perturbations.
 
\acknowledgements
S.K. was supported by the Korea Research Foundation Grant
 funded by the Korean Government(MOEHRD)(KRF-2006-214-C00013).
This work was supported in part by an NSERC Discovery Grant to
R.B., by a FQRNT Team Grant, and by funds from the Canada Research
Chair program.

\end{document}